\documentclass[12pt]{article}
\usepackage{amsmath,amssymb}
\usepackage{latexsym,graphicx,epsfig,wrapfig}

\newcommand{\be}{\begin{equation}}
\newcommand{\ee}{\end{equation}}
\newcommand{\bes}{\begin{subequations}}
\newcommand{\ees}{\end{subequations}}
\newcommand{\bea}{\begin{eqnarray}}
\newcommand{\eea}{\end{eqnarray}}
\newcommand{\ba}{$\begin{array}}

\newcommand{\ea}{\end{array}$}

\newcommand{\bear}{\begin{equation}\begin{array}}

\newcommand{\bm}{\boldmath}
\newcommand{\fr}[2]{\dfrac{{ #1}}{{ #2}}}

\newcommand{\la}{\langle}
\newcommand{\ra}{\rangle}

\newcommand{\fn}[1]{\footnote{{\sf #1}}}
\def\vak{{\varkappa}}

\newsavebox{\fmbox}

{\end{list}}
\newcounter{enumct}

\allowdisplaybreaks 

\title {\bf \bm Tree-level unitarity constraints
in the 2HDM with $CP$-violation}

\author{\em I.F. Ginzburg\thanks{email: ginzburg@math.nsc.ru},
I.P. Ivanov\thanks{email: igivanov@mail.desy.de}\\
{\small Sobolev Institute of Mathematics, Novosibirsk, Russia}}

\begin{document}

\maketitle

\abstract{ We obtain tree-level unitarity constraints for the Two
Higgs Doublet Model (2HDM) with explicit $CP$-violation. We limit
ourselves to the case with soft violation of the discrete $Z_2$
symmetry of theory. The key role is played by the rephasing
invariance of the 2HDM lagrangian. Our simple approach for
derivation of these constraints can be easily transferred to other
forms of Higgs sector. We briefly discuss correspondence between
possible violation of tree level unitarity limitation and physical
content of theory.} \vspace{1cm}

The Electroweak Symmetry Breaking in the Standard Model is
described usually with the Higgs mechanism. In its simplest
variant, an initial Higgs field is an isodoublet of scalar fields
with weak isospin $\vec{\sigma}$. The simplest extension of the
Higgs sector known as two-Higgs-doublet model (2HDM) consists in
introducing two Higgs weak isodoublets of scalar fields $\phi_1$
and $\phi_2$ with hypercharge $Y=+1$ (for a review, see \cite{Hunter}).

We consider the Higgs potential of 2HDM in the form
 \bear{c}
V=\dfrac{\lambda_1}{2}(\phi_1^\dagger\phi_1)^2
+\dfrac{\lambda_2}{2}(\phi_2^\dagger\phi_2)^2
+\lambda_3(\phi_1^\dagger\phi_1) (\phi_2^\dagger\phi_2)
+\lambda_4(\phi_1^\dagger\phi_2) (\phi_2^\dagger\phi_1) \\
+\dfrac{1}{2}\left[\lambda_5(\phi_1^\dagger\phi_2)^2+
\lambda_5^*(\phi_2^\dagger\phi_1)^2\right] \\
-\dfrac{1}{2}\left\{m_{11}^2(\phi_1^\dagger\phi_1)
+\left[m_{12}^2 (\phi_1^\dagger\phi_2)+ (m_{12}^2)^*
(\phi_2^\dagger\phi_1) \right]
+m_{22}^2(\phi_2^\dagger\phi_2)\right\}\,.
 \end{array}\label{higgspotential}
 \end{equation}
Here, $\lambda_{1-4}$, $m_{11}^2$ and $m_{22}^2$ are real
(due to hermiticity of the potential), while $\lambda_{5}$
and $m_{12}$ are, in general, complex.

This potential with real coefficients describes the theory
without CP violation in the Higgs sector while complex
values of some coefficients here make CP violation in Higgs
sector possible (a more detailed discussion of many points
here and references see in ref.~\cite{futurereview}).

$\bullet$ The Higgs--Higgs scattering matrix at high enough energy
at tree level contain only $s$--wave amplitudes; it is described
by the quartic part of this potential only. The {\it tree level
unitarity constraints} require that the eigenvalues of this
scattering matrix be less than the unitarity limit.

Since the coefficients of the scattering matrix at high enough
energy are given only by parameters $\lambda_i$ of the Higgs
potential, the tree-level unitarity constraints are written as
limitations on parameters $\lambda_i$ (for example, in the minimal
SM, with one Higgs doublet and
$V=(\lambda/2)(\phi^\dagger\phi-v^2/2)^2$, such unitarity
constraint looks as $\lambda<8\pi$).

Until now, these constraints were considered only for the case
without CP violation, i.e. with all coefficients of potential
(\ref{higgspotential}) real, \cite{Akeroyd:2000wc}. Below  we
extend these results to the case with explicit CP violation and
rederive these constraints in a transparent way, suitable for
other extended Higgs sectors discussed in literature.

$\bullet$ The crucial role in the 2HDM is played by the
discrete $Z_2$-symmetry, i.e. symmetry under transformation
\begin{equation}        \label{Eq:Z2-symmetry}
(\phi_1 \leftrightarrow\phi_1,\phi_2
\leftrightarrow-\phi_2) \quad \text{or} \quad (\phi_1
\leftrightarrow-\phi_1, \phi_2\leftrightarrow\phi_2).
\end{equation}
This symmetry forbids ($\phi_1,\,\phi_2$) mixing.

With this symmetry, the CP violation in the Higgs sector is
forbidden and Flavor Changing Neutral Currents (FCNC) are
unnatural. In the "realistic"\ \ theory this $Z_2$ symmetry
is violated.

Potential (\ref{higgspotential}) contains the $m_{12}^2$ term, of
dimension two, which softly violates the $Z_2$ symmetry. Soft
violation implies that $Z_2$ symmetry is broken near the mass
shell, and is restored at small distances $\ll 1/M_i$, where $M_i$
are masses of Higgs particles\fn{ Some authors consider the "most
general" Higgs potential with also operators of dimension four,
$\left(\lambda_6(\phi_1^\dagger\phi_1)+
\lambda_7(\phi_2^\dagger\phi_2)\right) (\phi_1^\dagger\phi_2)
+h.c.$ which describe hard violation of $Z_2$ symmetry at all
distances. Unfortunately, these discussions are incomplete since
these potential terms should be supplemented  {\it for
renormalizability} by the mixed kinetic term like $\vak (D_{\mu}
\phi_1 )^{ \dagger}(D^{\mu} \phi_2)+h.c.$ (for more detailed
comments see \cite{futurereview}). We do not consider this case.}.

$\bullet$ Potential (\ref{higgspotential}) (and therefore the
entire lagrangian) is invariant under the global phase rotations
of both doublets $\phi_i\to \phi_ie^{-i\rho_0}$ with common phase
$\rho_0$ ({\it overall phase freedom}). Besides, the same physical
reality (the same set of observables) can be described by a class
of lagrangians that differ from each other by independent phase
rotation for each doublet \cite{Branco},~\cite{futurereview}
 \bes\label{rotations}
 \be
\phi_i\to e^{-i\rho_i}\phi_i,\quad \text{$\rho_i$
real}\quad (i=1,2),\quad 
\rho=\rho_2-\rho_1\,,
 \ee
accompanied by compensating phase rotations of parameters
of lagrangian:
 \begin{equation}
\lambda_{1-4}\to \lambda_{1-4}\,,\quad
m_{11(22)}^2\to m_{11(22)}^2\,,\ \
\lambda_5\to\lambda_5\, e^{-2i\rho}\,,\;\;
m_{12}^2\to m_{12}^2e^{-i\rho}\,.
\label{formtrans}
 \end{equation}
 \ees
The invariance of the physical picture in respect to this
transformation is called as {\it the rephasing invariance} and the
set of these physically equivalent lagrangians we call as the {\it
rephasing equivalent family}. This one-parametric family is
governed by phase difference $\rho$ which we call {\it the
rephasing gauge parameter}. Let us underline that this parameter
cannot be determined from any measurement, its choice is only a
matter of convenience.

This rephasing invariance is extended to the entire system of
fermions and scalars with hard violation of $Z_2$ symmetry if one
supplements transformations (\ref{rotations}) by similar
transformations for the hard $Z_2$ symmetry violating terms and
Yukawa terms (phases of fermion fields and Yukawa couplings).

$\bullet$ The doublets $\phi_i$ contain fields with weak isospin
projections $\sigma_z=\pm 1/2$:
 \be
 \phi_i=\left(\begin{array}{c}|+1/2\ra\\[1mm]
 |-1/2\ra\end{array}\right)\equiv
 \left(\begin{array}{c}\phi_i^+\\[1mm]
 n_i+\fr{v_i}{\sqrt{2}}\end{array}\right),\;\;
 n_i=\fr{\eta_i+i\xi_i}{\sqrt{2}}\;\;\; (i=1,\,2)\,,\; \begin{array}{c}
 v_1=v\cos\beta\\[1mm]
 v_2=v\sin\beta\end{array}\, .\label{fields}
 \ee
Here $v_i\equiv \la \phi_i\ra$ are the vacuum expectation values
(v.e.v.'s) of $\phi_i$, which are, in general, complex.
For the conjugate fields the isospin projections are $\phi^-
=|-1/2\ra$, $n^*=-|+1/2\ra$.

Adjusting the global phase, one can make one of these
v.e.v.'s (e.g. $v_1$) real. The rephasing transformation
mixes $\eta_i$ and $\xi_i$ and change phase of $v_2$.

Two complex isodoublet fields have eight degrees of
freedom. Three of them correspond to Goldstone fields
$\phi^\pm_1\cos\beta +\phi^\pm_2\sin\beta$, $\xi_1\cos\beta
+\xi_2\sin\beta$ (which are transformed to longitudinal
components of gauge bosons $W_L$, $Z_L$). The combinations
$H^\pm= \phi^\pm_2\cos\beta -\phi^\pm_1\sin\beta$ describe
the observable charged Higgs bosons. The scalar
$\eta_1,\,\eta_2$ and pseudoscalar fields $A=\xi_2\cos\beta
-\xi_1\sin\beta$ mix to the observable neutral Higgs bosons
$h_1$, $h_2$, $h_3$ (which might have no definite CP
parity).

$\bullet$ A natural way for derivation of the tree level unitarity
constraints is to construct the scattering matrix for all the
physical Higgs--Higgs (as well as $Z_L h_i$, $W_L W_L$, etc.)
states in the tree approximation at high enough energy (where
threshold effects are inessential) and diagonalize it. This very
way was realized in the first derivation of such constraint in the
Minimal Standard Model \cite{LQT}.

The tree-level unitarity constraints are written for the
scattering matrix as the limitations for its eigenvalues. They can
be obtained in any basis related to the physical basis by a
unitarity transformation. For the considered problem, derivation
simplifies in the basis of the non--physical Higgs fields
$\phi_i^\pm$, $\eta_i$, $\xi_i$, \cite{Akeroyd:2000wc,kanemura93}.
The calculations of the mentioned works were limited to the case
of CP--conserving Higgs sector, with all $\lambda_i$ real.

The following simple observation allows us to extend these results
to a more general case with CP violation. Let us repeat that the
unitarity constraints are written for the very high energy
Higgs--Higgs scattering matrix, which is expressed via quartic
terms of potential $\lambda_i$ only. {\it Starting from
representation with complex $\lambda_5$, one can perform rephasing
transformation (\ref{rotations})} (with $\rho=-arg(\lambda_5)/2$)
{\it to obtain the real parameter $\lambda_{5r}$ and derive
unitarity constraints just as in the CP-conserving case. Since the
physical picture is invariant under such rephasing transformation,
the results with $\lambda_{5r}=|\lambda_5|$ correspond also to the
initial situation}\fn{ This rephasing gauge is different from that
which is useful for describing CP-violation \cite{futurereview}.}.

$\bullet$ With our choice of rephasing representation with
real $\lambda_5$, the CP is conserved at very high energy,
so it is sufficient to consider complex neutral fields
$n_i$ (\ref{fields}) instead of separate scalars $\eta_i$
and pseudoscalars $\xi_i$ (which would mix in an arbitrary
rephasing gauge). In the the high-energy scalar-scalar
scattering, the total weak isospin $\vec\sigma$, the total
hypercharge $Y$ are conserved. Besides, since the quartic
part of the Higgs potential (\ref{higgspotential})
conserves $Z_2$ symmetry, the Higgs--Higgs states can also
be classified according to their value of the $Z_2$-parity:
$\phi_i\phi_i$, $\phi_i\phi_i^*$, etc. will be called the
$Z_2$-even states, and $\phi_1\phi_2$, $\phi_1\phi_2^*$,
etc. will be called the $Z_2$-odd states, and this $Z_2$
parity is also conserved at very high energy.

Table 1 shows the classification of the two-scalar states
constructed from fields $\phi_i^\pm$, $n_i$ and $n_i^{*}$
according to the $Z_2$ parity, hypercharge, weak isospin and its
$z$-projection. Only the transitions between the states within
each cell of Table 1 are possible\fn{ The double-charged states
with $Y=2$, $\sigma=\sigma_z=1$ were omitted in the analysis of
\cite{Akeroyd:2000wc}.}.

\begin{center}
\begin{tabular}{|c|c||c||c|||c||c||}\hline
&&\multicolumn{2}{|c||}{$Y=2$}
&\multicolumn{2}{|c||}{$Y=0$}\\\hline $\sigma$&$\sigma_z$&
$Z_2$ even&$Z_2$ odd & $Z_2$ even&$Z_2$ odd
\\\hline\hline
    &1&$\left(\begin{array}{c}
 \phi_1^+\phi_1^+ \\
 \phi_2^+\phi_2^+
\end{array}\right)$&$\phi_1^+\phi_2^+$&
 $\left(\begin{array}{c}
 \phi_1^+n_1^*\\
  \phi_2^+n_2^*
\end{array}\right)$&
$\left(\begin{array}{c}
 \phi_1^+n_2^*\\
  \phi_2^+n_1^*
\end{array}\right)$\\\cline{2-6}
1&0&$\left(\begin{array}{c}
 \phi_1^+n_1 \\
 \phi_2^+n_2
\end{array}\right)$&
$\fr{\phi_1^+n_2+\phi_2^+n_1}{\sqrt{2}}$&
 $\left(\begin{array}{c}
 \fr{\phi_1^+\phi_1^- -n_1n_1^*}{\sqrt{2}}\\
  \fr{\phi_2^+\phi_2^- -n_2n_2^*}{\sqrt{2}}
\end{array}\right)$&
$\left(\begin{array}{c}
 \fr{\phi_1^+\phi_2^- -n_1n_2^*}{\sqrt{2}}\\
  \fr{\phi_2^+\phi_1^- -n_2n_1^*}{\sqrt{2}}
\end{array}\right)$ \\\cline{2-6}
&-1&$\left(\begin{array}{c}
 n_1n_1 \\
 n_2n_2
\end{array}\right)$&$n_1n_2$
& $\left(\begin{array}{c}
 \phi_1^-n_1\\
  \phi_2^-n_2
\end{array}\right)$&
$\left(\begin{array}{c}
 \phi_1^-n_2\\
  \phi_2^-n_1
\end{array}\right)$
\\\hline
0&0&absent& $\fr{\phi_1^+n_2-\phi_2^+n_1}{\sqrt{2}}$
 &$\left(\begin{array}{c}
 \fr{\phi_1^+\phi_1^- +n_1n_1^*}{\sqrt{2}}\\
  \fr{\phi_2^+\phi_2^- +n_2n_2^*}{\sqrt{2}}
\end{array}\right)$&
$\left(\begin{array}{c}
 \fr{\phi_1^+\phi_2^- +n_1n_2^*}{\sqrt{2}}\\
  \fr{\phi_2^+\phi_1^- +n_2n_1^*}{\sqrt{2}}
\end{array}\right)$\\ \hline\hline
\end{tabular}\\
\vspace{5mm} {\it Table 1. The two-Higgs states with
different quantum numbers.}
\end{center}
The states with $Y=-2$, $\sigma=1$ are obtained from those for
$Y=2$ by charge conjugation and change of sign of $\sigma_z$. For
$Y=\pm 2$, the $Z_2$ even state with $\sigma=0$ cannot occur due
to Bose-Einstein symmetry of identical scalars.

The scattering matrices for different states are determined
completely by values of the hypercharge, the total weak isospin
and the $Z_2$ parity of the initial Higgs--Higgs state, they are
independent of $\sigma_z$. The scattering matrices for each set of
these quantum numbers are listed in Table 2, their coefficients
are easily seen from the potential (\ref{higgspotential}).
\begin{center}
\begin{tabular}{|c|c||c||c||}\hline
 $Y$&$\sigma$& $Z_2$ even&$Z_2$ odd\\ \hline
$\pm 2$&1&$\left(\begin{array}{cc}
 \lambda_1 & \lambda_{5r}\\
 \lambda_{5r}&  \lambda_2
\end{array}\right)$&$\lambda_3+\lambda_4$\\\hline
$\pm 2$&0&--&$\lambda_3-\lambda_4$\\\hline\hline
 0&1&
 $\left(\begin{array}{cc}
 \lambda_1 & \lambda_4\\
 \lambda_4&  \lambda_2
\end{array}\right)$&$\left(\begin{array}{cc}
 \lambda_3 & \lambda_{5r}\\
 \lambda_{5r}&  \lambda_3
\end{array}\right)$\\\hline
0&0&$\left(\begin{array}{cc}
 3\lambda_1&2\lambda_3+\lambda_4\\
  2\lambda_3+\lambda_4& 3\lambda_2
\end{array}\right)$
& $\left(\begin{array}{cc}
\lambda_3+2\lambda_4& 3\lambda_{5r}\\
 3\lambda_{5r}&\lambda_3+2\lambda_4
 \end{array}\right)$\\\hline\hline
\end{tabular}\\\vspace{5mm}
\noindent{\it Table 2. Scattering matrices for different
Higgs--Higgs states (with factor $1/(8\pi)$). }
 \end{center}
The case $Y=0$, $\sigma_z=0$ demands special discussion. Let us
consider, for example, term
$V_1=\lambda_1(\phi_1^\dagger\phi_1)^2/2$ in the potential. In
terms of operators $\hat{\phi}^\pm$, $\hat{n}$ (omitting subscript
1 for brevity), we have $\hat{V}_1=\lambda_1(\hat\phi^-\hat\phi^+
+ \hat n \hat n^*)^2/2$. Within the subspace of neutral
two-particle states with $Y=0$, one can rewrite $\hat{V}_1$ after
simple combinatorics as
 \bea
&&\hat{V}_1 ={\lambda_1\over 2} \left[4
|\phi^+\phi^-\rangle\langle \phi^+\phi^-| + 2
|\phi^+\phi^-\rangle\langle nn^*| + 2 |nn^*\rangle \langle
\phi^+\phi^-|
+ 4 |nn^*\rangle \langle nn^*|\right]\nonumber\\[3mm]
&&=3\lambda_1 \dfrac{|\phi^+\phi^-\rangle+|n
n^*\ra}{\sqrt{2}}\cdot\dfrac{\la\phi^+\phi^-|+\la n
n^*|}{\sqrt{2}} + \lambda_1\dfrac{|\phi^+\phi^-\rangle-|n
n^*\ra}{\sqrt{2}}\cdot\dfrac{\la\phi^+\phi^-|-\la n
n^*|}{\sqrt{2}}\,. \nonumber
 \eea
One can see precisely these numerical coefficients in
Table 2.

The classification scheme based on the quantum numbers of EW
theory $\sigma$, $\sigma_z$, $Y$, and $Z_2$ looks more {\em
natural} for the considered problems than both the
$O(4)$-classification (in the minimal SM) introduced in \cite{LQT}
and the scheme based on new quantum numbers $C$, $G$, and $Y_\pi$
introduced in \cite{kanemura93}.

The scheme proposed above can be readily exploited in the study of
some other multi-Higgs sectors. For example, the analysis in the
cases of widely discussed 2 doublet + singlet model or
doublet--triplet model should be also very simple, the analysis of
three-doublet Higgs model, etc, is expected to be not very
difficult.

Note that $\lambda_{5r}\equiv |\lambda_5|$. Therefore, one
can present all eigenvalues of these scattering matrices
$\Lambda^{Z_2}_{Y\sigma\pm}$ even for complex values of
$\lambda_5$ as
 \bea
 \Lambda^{even}_{21\pm}&=&\fr{1}{2}\left(\lambda_1+\lambda_2\pm
 \sqrt{(\lambda_1-\lambda_2)^2+4|\lambda_5|^2\,}\;\right),
\quad \Lambda^{odd}_{21}=\lambda_3+\lambda_4\,,\quad
\Lambda^{odd}_{20}=\lambda_3-\lambda_4\,,\nonumber\\
\Lambda^{even}_{01\pm}&=&\fr{1}{2}\left(\lambda_1+\lambda_2\pm
 \sqrt{(\lambda_1-\lambda_2)^2+4\lambda_4^2\,}\;\right),
 \quad \Lambda^{odd}_{01\pm}=\lambda_3\pm|\lambda_5|\,,
 \label{eigen}\\
\Lambda^{even}_{00\pm}&=&\fr{1}{2}\left[3(\lambda_1+\lambda_2)\pm
 \sqrt{9(\lambda_1-\lambda_2)^2+4(2\lambda_3+\lambda_4)^2\,}\;\right),
 \ \ \Lambda^{odd}_{00\pm}=\lambda_3+2\lambda_4\pm
 3|\lambda_5|\,.\nonumber
\eea

The tree-level unitarity constraints can be written now as
 \be
|\Lambda^{Z_2}_{Y\sigma\pm}|<8\pi\,.\label{unitconstr}
 \ee
They differ from those obtained in
ref.~\cite{Akeroyd:2000wc} only by the change $\lambda_5
\to|\lambda_5|$.

$\bullet$ Note that in the considered tree approximation the
masses of Higgs bosons are composed from quantities $\lambda_i $
and quantity $\nu \propto Re(\overline{m}_{12}^2)/v_1v_2$ (the
quantity $m_{12}^2$ in a special rephasing gauge different from
that used above -- see \cite{futurereview} and \cite{masses} for
details). Since parameter $m_{12}^2$ does not enter the quartic
interactions, the above unitarity constraints, generally, do not
set any limitation on masses of observable Higgs bosons, which was
explicitly noted in \cite{kanemura93}. Reasonable limitations on
these masses can be obtained for some specific values of $\nu$.
For example, for reasonably small value of $\nu$ one can have the
lightest Higgs boson mass of about 120 GeV and the masses of other
Higgs bosons can be up to about 600 GeV without violation of
tree-level unitarity \cite{futurereview}. At large $\nu$, masses
of all Higgs bosons except the lightest one can be very large
without violation of unitarity constraint.

$\bullet$ Let us discuss briefly some new features, which are
brought up by the situation with unitarity constraints in 2HDM.

The unitarity constraints were obtained first \cite{LQT} in the
minimal SM. In this model, the Higgs boson mass
$M_H=v\sqrt{\lambda}$, and its width $\Gamma$ (given mainly by
decay to longitudinal components of gauge bosons $W_L$, $Z_L$)
grows as $\Gamma\propto M_H^3$. The unitarity limit corresponds to
the case when $\Gamma_H\approx M_H$, so that the physical Higgs
boson disappears, the strong interaction in the Higgs sector is
realized as strong interaction of longitudinal components of gauge
bosons $W_L$, $Z_L$ at $\sqrt{s}> v\sqrt{\lambda}\gtrsim
v\sqrt{8\pi}\approx 1.2$ TeV. Therefore, if $\lambda$ exceeds the
tree-level unitarity limitation, the discussion in terms of
observable Higgs particle becomes meaningless, and a new physical
picture for the Electroweak Symmetry Breaking in SM arises.

Such type of correspondence among the tree level unitarity limit,
realization of the Higgs field as more or less narrow resonance
and a possible strong $W_LW_L$ and $Z_LZ_L$ interaction, can
generally be violated in the 2HDM if values of $\lambda_i$ differ
from each other essentially. Large number of degrees of freedom of
2HDM generates situations when some of Higgs bosons of this theory
are "normal"\ more or less narrow scalars (whose properties can be
estimated perturbatively), while the other scalars and (or) $W_L$,
$Z_L$ interact strongly at sufficiently high energy. It can happen
that some of the latter can be realized as physical particles,
while the other disappear from particle spectrum like Higgs boson
in SM with large $\lambda$. In such cases the unitarity
constraints work in different way for different {\it physical}
channels. The list of
possibilities will be studied elsewhere.\\

We are thankful to M. Krawczyk and V.G.~Serbo for valuable
comments. This work was supported by INTAS grants 00-00679
and 00-00366, RFBR grant 02-02-17884, NSh-2339.2003.2 and
grant ``Universities of Russia".

\end{document}